# Adaptive Energy Management for Real Driving Conditions via Transfer Reinforcement Learning

Teng Liu, *Member*, *IEEE*, Wenhao Tan, Xiaolin Tang, Jiaxin Chen, Dongpu Cao

*Abstract*—This article proposes a transfer reinforcement learning (RL) based adaptive energy managing approach for a hybrid electric vehicle (HEV) with parallel topology. This approach is bi-level. The up-level characterizes how to transform the Q-value tables in the RL framework via driving cycle transformation (DCT). Especially, transition probability matrices (TPMs) of power request are computed for different cycles, and induced matrix norm (IMN) is employed as a critical criterion to identify the transformation differences and to determine the alteration of the control strategy. The lower-level determines how to set the corresponding control strategies with the transformed Q-value tables and TPMs by using model-free reinforcement learning (RL) algorithm. Numerical tests illustrate that the transferred performance can be tuned by IMN value and the transfer RL controller could receive higher fuel economy. The comparison demonstrates that the proposed strategy exceeds the conventional RL approach in both calculation speed and control performance.

## I. INTRODUCTION

HYBIRD electric vehicles (HEVs) show great potential to promote fuel economy and decrease air pollution in recent decades [1]. To optimize a pre-selected cost function (e.g., fuel consumption, running cost, and harmful emissions), HEVs designers need to set appropriate energy management for multiple power sources [2]. One difficult problem needs to be overcame is how to adapt to different driving conditions. Many researchers address this problem by studying the driving cycles of vehicles.

Driving cycle is an indication of vehicle speed versus time, which is often applied to capture driver behaviors versus traffic situations [3], [4]. Different driving cycles and overall fuel consumption properties of vehicles usually lead to different energy management strategies of HEVs. As a result, an optimal energy management strategy for a special driving cycle may become sub-optimal for another driving cycle [5]. So, researchers show increasing interests in designing adaptive energy management strategy to accommodate different driving cycles that can be met in practice.

Methods used to adapt various driving cycles include primarily two types: model-based ones and model-free ones. Model-based approaches have been well studied in the recent two decades. For example, a microtrip approach [6] or Markov chain (MC)-based technique [7] is proposed to generate the new driving cycles to design adaptive control. Furthermore, a novel approach is applied to construct a new driving cycle with a speed-acceleration frequency distribution plot and a quasi-random selection mechanism in [8] and [9]. Lee *et al.* proposed another MC-based generator to select speed and acceleration as states and extract naturalistic information as transition probability matrices [10]. In [11], Kruse *et al.* scaled the velocity and time in the Urban Dynamometer Driving Schedule (UDDS) by the factors ranging from 1.1 to 1.4, which leads to the increase of the acceleration and mean velocity and keeps the driving distance the same.

Tests indicate that the adaptation of driving cycles generation-based energy management strategy is heavily influenced by the accuracy of generation algorithms. So, various optimization techniques are utilized to search adaptive control in one or multiple driving cycles for HEVs. When full driving cycle is given in advance, deterministic dynamic programming (DDP) [12], [13], Pontryagin's minimum principle (PMP) [14], and the equivalent consumption minimization strategy (ECMS) [15], [16] have been developed to get the global optimal control decision for power split in HEVs.

However, these pre-computed controls do not fit, when the full driving cycle is not given [17]. As alternatives, model predictive control (MPC) and stochastic dynamic programming (SDP) were studied during recent years. Peng *et al.* proposed a stochastic control strategy based on the Markov decision process (MDP) for a parallel hybrid electric [18], [19]. In [20], Vagg *et al.* addressed the robustness characteristics of the SDP method and aimed to improve battery health and lessen the motor temperature via incorporating the square of battery charge. In addition, Romaus *et al.* added the impacts of driver and traffic into the energy management problem and solved this

The work was supported in part by National Natural Science Foundation of China (51705044). (Corresponding authors: Xiaolin Tang)

T. Liu is with Department of Automotive Engineering and the State Key Laboratory of Mechanical Transmission, Chongqing University, Chongqing 400044, China, and also with Mechanical and Mechatronics Engineering Department, University of Waterloo, N2L 3G1, Canada. (email: tengliu17@gmail.com)

W. Tan, X. Tang, and J. Chen are with State Key Laboratory of Mechanical Transmissions, College of Automotive Engineering, Chongqing University, Chongqing, 400044, PR China. (email: tangxl0923@cqu.edu.cn)

D. Cao is with Mechanical and Mechatronics Engineering Department, Waterloo University, N2L 3G1, Canada. (email: hong.wang@uwaterloo.ca)

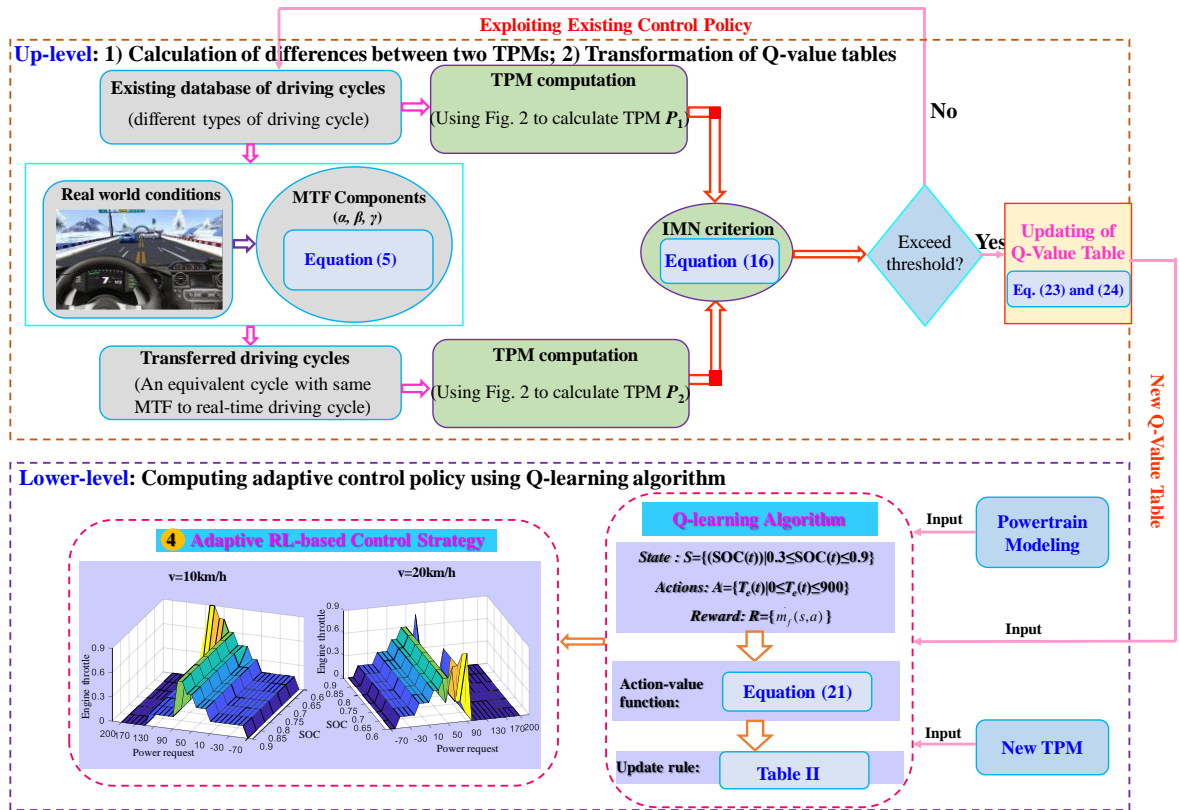

Fig. 1. The bi-level approach for adaptive and model-free energy management strategy.

problem using SDP to realize online control [21]. In MPC-based optimization, Borhan *et al.* considered closed-loop modeling of series HEV and applied MPC to promoted its online performance [22]. Furthermore, Zeng *et al.* constructed a stochastic MPC controls via synthesizing the area information of hilly regions together, such as terrain, vehicle location and traveling direction [23]. Nevertheless, the prediction accuracy affects the control performance of the MPC method to a great extent [24].

The aforementioned processes all need accurate vehicle modeling. It brings considerable model parameter calibration cost. To solve this problem, researchers began to show interests in model-free methods. Reinforcement learning (RL) is introduced as an alternative to search the adaptive control for HEVs by defining appropriate reward function in advance [25, 26]. For instance, in [27], Liu *et al.* systematically discussed the optimality and the learning ability of the RL approach, and they simultaneously proposed a real-time control strategy via power request recursive algorithm and a certain definitive criterion in [28]. However, it must indicate that fuel consumption may even increase if the deduced controls do not match the present driving cycles [29]. Especially, when the vehicle speed change sharply (e.g., from the highway to the urban cycle), the control strategy in [27] and [28] cannot handle the new driving conditions. Thus, Guzzella *et al.* measured the driving cycles equivalently with the integral values named mean tractive force (MTF) in [30], and Nyberg *et al.* applied the MTF components to achieve the equivalent transformation of driving cycles and motivate the vehicle in similar driving conditions [31].

Recent studies reveal that the memory matrix transformation in reinforcement learning (RL) framework could be an efficient and useful tool for adaptive and real-time power split optimization [32]. Memory matrix represents a memory of the corresponding state-action pair, i.e., Q-value tables in RL, which can be updated to adapt to different driving cycles. Hence, this paper proposes a transfer RL-enabled adaptive and model-free energy management strategy to improve the fuel efficiency of a parallel HEV online.

As shown in Fig. 1, the proposed approach is bi-level. Three potential contributions emerge in this article: (1) the up-level characterizes how to transfer memory matrix through driving cycles transformation (DCT). Before transformation, the existing driving cycle is called primitive cycle, and after transformation, the resulted cycle is referred to as transferred cycle; (2) Induced matrix norm (IMN) is employed as a critical criterion to identify the differences of various memory matrices and to determine the alteration of control strategy due to its quicker calculational speed; (3) the lower-level determines how to set the corresponding control strategies based on the transferred Q-value tables.

The computational flowchart of the bi-level approach is described as follow: the inputs are the real-world driving cycles, and the corresponding MTF components are calculated first. Based on the MTF, the historical driving cycles database is transformed into an equivalent one. Then, the transition probability matrices (TPMs) related to these two cycles are calculated and differences between them are decided by IMN. If this IMN value exceeds the threshold value, the memory matrix (Q-value table) in RL needs to be updated by transformation, and the relevant controls will be computed. The new controls are feasible for the current driving cycles. When the driving cycles change again, the above process repeats to design new controls.

The transferred performance is first discussed by numerical tests. Then, the optimality of this new approach is evaluated by comparing with the dynamic programming (DP) algorithm, and the adaptability is demonstrated through comparing with the conventional RL method in fuel economy. The conventional RL method means the memory matrix does not change along with the driving conditions. Results show that the presented strategy is superior to the conventional RL technique in calculation speed and control performance. These advantages make it feasible to apply the proposed control in real-world driving conditions.

The following paper is described as follows. The DCT and IMN of the up-level are given in Section II. Section III formulates the optimization control problem and introduces the transfer process of memory matrices. Simulation results related to the transformation and comparison are depicted in Section IV. Finally, Section V summaries the conclusions.

## II. THE UP-LEVEL

Driving cycle transformation process is illustrated in this section. It can be formulated as a non-linear optimization problem. First, MTF components are defined to represent the statistical characteristics of the driving cycles. Based on them, constraints for transformation are introduced. Then, vehicle driveability is taken as a cost function to decide the optimal transferred driving cycle. Finally, IMN is applied to evaluate the availability of transformation by comparing the differences in TPMs of power demand.

### A. MTF Components

The MTF is defined as the tractive energy divided by the distance traveled for a whole driving cycle, which is integrated over the entire time interval $T=[t_0, t_f]$ as follow

$$\bar{F} = \frac{1}{x_L} \int_{t_0}^{t_f} F(t)v(t)dt \tag{1}$$

where $x_L$ is the total distance traveled in a certain driving cycle and depicted as $\int v(t)dt$. $v$ means the speed in regard to a certain driving cycle. $F$ is the longitudinal force to propel the vehicle, which is computed as

$$\begin{cases} F = F_a + F_f + F_m \\ F_a = \frac{1}{2}\rho_a C_d A v^2, F_f = mg \cdot f, F_m = ma \end{cases} \tag{2}$$

where $F_a$ is the aerodynamic drag, $F_f$ is the rolling resistance and $F_m$ is the inertial force. $\rho_a$ is the air density, $C_d$ is the aerodynamic coefficient, and $A$ is the fronted area. $m$ is the curb weight, $g$ is the gravitational constant, $f$ is the rolling friction coefficient and $a$ is the acceleration. According to the force $F$ imposed on the vehicle powertrain, the operation modes of powertrain are divided as coasting, traction and braking [30].

Together with the idling operation, the entire time interval $T$ is denoted as

$$\begin{cases} T = T_{tr} \cup T_{co} \cup T_{br} \cup T_{id} \\ T_{tr} = \{t|F(t) > 0, v(t) \neq 0\}, T_{co} = \{t|F(t) = 0, v(t) \neq 0\} \\ T_{br} = \{t|F(t) < 0, v(t) \neq 0\}, T_{id} = \{t|v(t) = 0\} \end{cases} \tag{3}$$

where $T_{tr}$ and $T_{co}$ are the traction-mode and coasting-mode region, severally. $T_{br}$ represents the time interval of braking and $T_{id}$ is another one for idling.

Since the powertrain does not provide any positive forces in the coasting and braking regions, the traction regions are those when the powertrain provides positive power to the wheels. The MTF is rewritten as following

$$\bar{F} = \bar{F}_{tr} = \frac{1}{x_L} \int_{t \in T_{tr}} F(t)v(t)dt = \bar{F}_a + \bar{F}_f + \bar{F}_m \tag{4}$$

The MTF components ($\alpha$, $\beta$, $\gamma$) are statistic characteristics measures for a driving cycle that are defined as [31]

$$\begin{cases} \alpha(v(t)) = \bar{F}_a / \frac{1}{2}\rho_a C_d A = \frac{1}{x_L} \int_{t \in T_{tr}} v^3(t)dt \\ \beta(v(t)) = \bar{F}_f / mg \cdot f = \frac{1}{x_L} \int_{t \in T_{tr}} v(t)dt \\ \gamma(v(t)) = \bar{F}_m / m = \frac{1}{x_L} \int_{t \in T_{tr}} a \cdot v(t)dt \end{cases} \tag{5}$$

It is obvious that MTF components ($\alpha$, $\beta$, $\gamma$) are related to the speed and acceleration for a specific driving cycle. These measures are employed as the constraints for driving cycles transformation.

### B. Constraints for Transformation

To determine the equivalence of two driving cycles, the quantitative standards $\alpha$, $\beta$, and $\gamma$ are applied as objective parameters to restrain the equivalence transformation. To follow the practical experience, we assume that the time intervals for non-traction and traction are invariable after the transformation, which means $T_{tr}(\tilde{v}) = T_{tr}(v)$, wherein $v$ and $\tilde{v}$ indicate the primitive and transferred driving cycles, respectively. Thus, the vehicle speed cannot decrease or increase too much from one vehicle speed point to the next during transformation.

However, the vehicle speed within each region may be altered [33]. Based on the real-time and scarce driving cycle, the targets measures $\alpha'$, $\beta'$, and $\gamma'$ are first determined, and then the existent driving cycle database can be transferred into an equivalent one based on the vehicle excitation constraints as follow:

1) Restriction on $\alpha$: As the target $\alpha'$ is known a prior, the following constraint could be forced on the transferred driving cycle as:

$$\frac{1}{x_L} \int_{t \in T_{tr}} \tilde{v}^3(t)dt = \alpha' \tag{6}$$

$$\Rightarrow (discrete) \frac{\sum_{i \in T_{tr}} \tilde{v}_i^3 \Delta t}{\sum \tilde{v}_i \Delta t} - \alpha' = 0 \Rightarrow g_1(\tilde{v}, T_{tr}, \alpha') = 0$$

where $\triangle t$ means the sampling time and $x_L = \sum \tilde{v}_i \triangle t$. $i \in T_{tr}$ are the indices wherein the homologous vehicle speed points are in the traction region.

2) **Restriction on $\beta$**: For a specific measure $\beta'$, the transferred driving cycle should satisfy the constraint as follow:

$$\frac{1}{x_L}\int_{t\in T_{tr}} \tilde{v}(t)dt = \beta' \quad (7)$$

$$\Rightarrow (discrete)\frac{\sum_{i\in T_{tr}} \tilde{v}_i \Delta t}{\sum \tilde{v}_i \Delta t} - \beta' = 0 \Rightarrow g_2(\tilde{v}, T_{tr}, \beta') = 0$$

3) **Restriction on $\gamma$**: As the target $\gamma'$ is known a prior, the following equality needs to be fulfilled by the transferred driving cycle:

$$\frac{1}{x_L}\int_{t\in T_{tr}} \dot{\tilde{v}}(t)\cdot\tilde{v}(t)dt = \gamma'$$

$$\Leftrightarrow \frac{1}{x_L}\int_{t\in T_{tr}} \frac{1}{2}\cdot d\tilde{v}^2(t) - \gamma' = 0 \quad (8)$$

$$\Rightarrow (discrete)\frac{\sum_k^{N_{tr}}[\frac{\tilde{v}^2(t)}{2}]_{t_{k,0}}^{t_{k,f}}}{\sum \tilde{v}_i \Delta t} - \gamma' = 0 \Rightarrow g_3(\tilde{v}, T_{tr}, \gamma') = 0$$

where $N_{tr}$ is the amount of the traction regions for a particular driving cycle and $k$ is the index, $k=1, 2, \ldots N_{tr}$. $t_{k,0}$ and $t_{k,f}$ represent the initial and final time for each traction region, respectively. It is apparent that the aforementioned three equalities $g_1$, $g_2$, and $g_3$ illuminate the certain restrictions for the quantitative standards of the transferred driving cycle.

To characterize traction region, the coasting velocity $v_{coast}$ is computed through using $F(t)=0$ and $a(t)=dv(t)/dt$ as follow

$$0 = F_a + F_f + F_m$$

$$\Rightarrow \dot{v}_{coast}(t) = -k_1^2 v_{coast}^2(t) - k_2^2 \quad (9)$$

$$\Rightarrow v_{coast}(t_k) = \frac{k_2}{k_1}\tan(\arctan(\frac{k_1}{k_2}v(t_{k-1}))) - k_1 k_2 \cdot (t_k - t_{k-1})$$

where $k_1^2 = 1/(2m)\cdot\rho_a C_d A$ and $k_2^2 = f\cdot g$. For general driving cycles in discrete time, the vehicle powertrain works as traction mode at $t_k$ means $v(t_k) > v_{coast}(t_k)$; and the powertrain operates in coasting mode when $v(t_k) = v_{coast}(t_k)$ or in braking mode if $v(t_k) < v_{coast}(t_k)$. Thus, the following inequality constraints need to be fulfilled for $t_i \in T_{tr}$

$$v_{coast}(t_i) < \tilde{v}(t_i) \quad t_i \in T_{tr}$$
$$\Rightarrow -\tilde{v}(t_i) + v_{coast}(t_i) < 0 \Rightarrow h_1(\tilde{v}, T_{tr}, v_{coast}) < 0 \quad (10)$$

Similarly, the work points of the non-traction mode result in an inequality described as follows

$$v_{coast}(t_j) \geq \tilde{v}(t_j) \quad t_j \notin T_{tr}$$
$$\Rightarrow -\tilde{v}(t_j) + v_{coast}(t_j) \geq 0 \Rightarrow h_2(\tilde{v}, T_{co}\cup T_{br}, v_{coast}) \geq 0 \quad (11)$$

### C. Cost Function for Transformation

As the MTF components are unique for a specific driving cycle, the equality constraints $g_1$, $g_2$, and $g_3$ and inequality constraints $h_1$ and $h_2$ can be employed to decide the transferred driving cycle. To choose an optimal equivalence driving cycle from a set of feasible solutions, the cost function that represents the drivability is minimized in this paper. Driving cycle transformation is equivalent to a non-linear program (NLP) constrained by equality and inequality equations as

$$\min_{\tilde{v}} f(\tilde{v}) = \left\|\frac{d^2\tilde{v}}{dt^2}\right\|_2^2$$
$$s.t. \quad g_i(\tilde{v}, T_{tr}, \alpha', \beta', \gamma') = 0, \, i = 1, 2, 3. \quad (12)$$
$$h_1(\tilde{v}, T_{tr}, v_{coast}) < 0$$
$$h_2(\tilde{v}, T_{co}\cup T_{br}, v_{coast}) \geq 0$$

where $d^2\tilde{v}/dt^2$ is the change rate of acceleration, which means the vehicle jerk to yield a smoothing driving cycle and improve the drivability. The sampling time $\Delta t$ is 1 second. The NLP on the driving cycle transformation has been solved by the *fmincon* function with an interior-point solver in MATLAB and it takes around one minute for each calculation.

### D. Induced Matrix Norm

Given a special HEV model, the power request with respect to a transferred driving cycle is supposed to be supplied by the engine and battery together as

$$\begin{cases} P_{req} = F\cdot v = P_e\cdot\eta_T + P_{bat}\cdot\eta_{mot}\cdot\eta_T \\ P_e = T_e\cdot\omega_e \end{cases} \quad (13)$$

where $P_e$ is the output power from the engine, $\eta_{mot}$ and $\eta_T$ are the efficiencies of the traction motor and transmission axle.

When the full driving cycle is not given, the driving power request can be modeled as a stationary Markov chain (MC). The transition probability of the power request is evaluated via maximum likelihood estimator (MLE) as

$$\begin{cases} p_{ik,j} = P(P_{req} = P_{req}^j | P_{req} = P_{req}^i, v = v_k) = \frac{N_{ik,j}}{N_{ik}} \\ N_{ik} = \sum_{j=1}^{M} N_{ik,j} \quad i, j = 1, 2, \ldots M \end{cases} \quad (14)$$

where $N_{ik,j}$ means the number for the transition from $P^i_{req}$ to $P^j_{req}$ that happened at vehicle speed $v_k$, $N_{ik}$ is the total transition counts initiated from $P^i_{req}$ at vehicle speed $v_k$, $k$ is the discrete time step, and $M$ is the amount of discrete power request index. Fig. 2 illustrates the computational diagram for the transition probability matrices (TPMs) of power request.

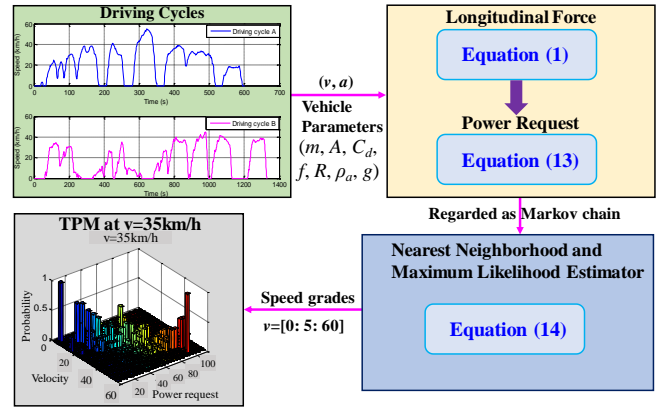

Fig. 2. Computational diagram of the TPM for power request.

For HEVs, different driving cycles match various controls. So, we need to update control strategies for different driving cycles. In order to distinguish the differences between original and transferred driving cycles for control alternation, we need to quantify the gap between two TPMs of power demand.

Assuming $P_1$ and $P_2$ are the primitive and transferred TPMs, respectively. The induced matrix norm (IMN) is introduced to quantify the similarity between them

$$IMN(P_1\|P_2) = \|P_1 - P_2\|_2 = \sup_{x \in R^M/\{0\}} \frac{|(P_1-P_2)x|}{|x|} \quad (15)$$

where *sup* depicts the supremum of a scalar, and $x$ is an $M \times 1$ dimension non-zero vector. The second-order norm in (15) can be reformulated as the following expression in online application [34]

$$IMN(P_1\|P_2) = \|P_1 - P_2\|_2 = \max_{1 \leq i \leq M} |\lambda_i(P_1 - P_2)| \\ = \max_{1 \leq i \leq M} \sqrt{\lambda_i((P_1-P_2)^T(P_1-P_2))} \quad (16)$$

where $P^T$ denotes the transpose of matrix $P$, and $\lambda_i(P)$ represents the eigenvalue of matrix $P$ for $i=1, \ldots, M$. Note that the closer the IMN is to zero, the more similar the TPM $P_1$ is to $P_2$. As the IMN value exceeds the threshold value, the memory matrix in RL needs to be updated. The transfer process of this matrix is designed in Section *III*.

## III. THE LOWER-LEVEL: POWERTRAIN MODEL AND REINFORCEMENT LEARNING

Energy management is formulated as a control optimization issue in this section. The definition of the optimization objective, control actions and state variables are given first. Furthermore, the updating process of the memory matrix in RL and the computation of the control strategy using Q-learning algorithm are illuminated [35], [36].

The schematic of the studied parallel powertrain configuration is sketched in Fig. 3. The maximum torque, power and speed of the motor are 600 Nm, 90 kW and 2400 rpm, respectively. The rated capacity of the battery pack is 60 Ah and its nominal voltage is 312.5 V. The maximum torque of the diesel engine is 900 Nm and its rated power is 155 kW. The main parameters of the powertrain are listed in Table I [37].

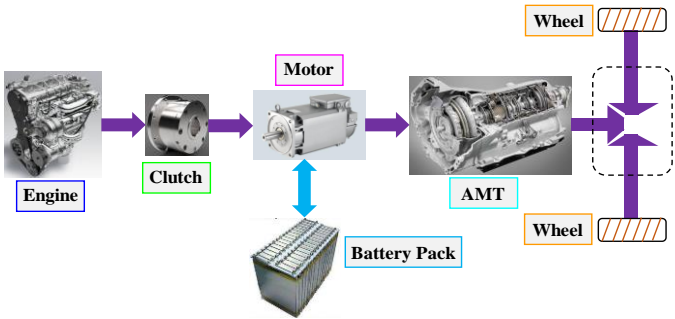

Fig. 3. A configuration of powertrain with parallel topology.

### A. Vehicle Powertrain Model

The objective of energy management for the parallel HEV is searching for optimal control under the constraints of components to minimize the cost function. In this article, the cost function is expressed by the sum of fuel consumption rate and charge sustenance over a finite horizon as

$$\begin{cases} J = \int_{t_0}^{t_f} [\dot{m}_f(t) + \sigma(\Delta SOC)^2]dt \\ \Delta SOC = \begin{cases} SOC(t) - SOC_{ref} & SOC(t) < SOC_{ref} \\ 0 & SOC(t) \geq SOC_{ref} \end{cases} \end{cases} \quad (17)$$

where $\sigma$ is a large positive weighting factor to restrict the terminal value of SOC ($\sigma$=10000 in this paper). The $SOC_{ref}$ means a pre-defined parameter to maintain the charge-sustaining constraints [38].

TABLE I
PARAMETERS OF MAIN COMPONENTS IN HEV

| Symbol | Items | Values |
|---|---|---|
| $m$ | Vehicle mass | 16000 kg |
| $A$ | Fronted area | 1.8 m² |
| $C_d$ | Aerodynamic coefficient | 0.55 |
| $\eta_T$ | Transmission axle efficiency | 0.9 |
| $\eta_{mot}$ | Efficiency of Traction motor | 0.95 |
| $f$ | Efficiency of Rolling resistance | 0.021 |
| $R$ | Radius of Tire | 0.508 m |
| $\rho_a$ | Air density | 1.293 kg/m³ |
| $g$ | Gravitational acceleration | 9.81 m/s² |

To guarantee the safety and reliability of the components, the optimization problem is subjected to the following inequality constraints

$$\begin{cases} T_{e,\min} \leq T_e(t) \leq T_{e,\max}, \quad \omega_{e,\min} \leq \omega_e(t) \leq \omega_{e,\max} \\ T_{m,\min} \leq T_m(t) \leq T_{m,\max}, \quad \omega_{m,\min} \leq \omega_m(t) \leq \omega_{m,\max} \\ P_{bat,\min} \leq P_{bat}(t) \leq P_{bat,\max}, \quad I_{bat,\min} \leq I_{bat} \leq I_{bat,\max} \\ SOC_{\min} \leq SOC(t) \leq SOC_{\max} \end{cases} \quad (18)$$

where $\omega_{e,\min}$, $\omega_{e,\max}$, $T_{e,\min}$, and $T_{e,\max}$ are the permitted lower and upper bounds of the engine speed and torque, respectively. $\omega_{m,\min}$, $\omega_{m,\max}$, $T_{m,\min}$, and $T_{m,\max}$ have the analogous meanings for the motor. $P_{bat,\min}$, and $P_{bat,\max}$ are thresholds of battery power admissible sets, same as the $SOC_{\min}$, $SOC_{\max}$, $I_{bat,\min}$ and $I_{bat,\max}$. Note that the paper does not consider the influences of battery aging and temperature.

Based on a quasi-static model [39], the fuel consumption rate and the total fuel consumption of the studied HEV in the entire time interval can be defined as follows

$$\begin{cases} \dot{m}_f = f(T_e, \omega_e) \\ fuel_{tot} = \int_{t_0}^{t_f} \dot{m}_f(t)dt \end{cases} \quad (19)$$

where $\omega_e$ and $T_e$ means the rotation speed and torque of the engine. The engine rotation speed is decided by the vehicle speed and transmission ratio. The engine torque is chosen as the control variable in this article.

The battery pack is modeled by a first-order internal resistance modeling, wherein the state of charge (SOC) of battery is selected as the state variable and developed as

$$\dot{SOC} = -I_{bat}(t)/Q_{bat}$$
$$\Rightarrow \dot{SOC} = -(V_{oc} - \sqrt{V_{oc}^2 - 4r_{in}P_{bat}})/2r_{in}Q_{bat} \quad (20)$$

where $I_{bat}$ means the battery current, $Q_{bat}$ represents the rated capacity and $P_{bat}$ denotes the output power. $V_{oc}$ is the open-circuit voltage and $r_{in}$ is the internal resistance.

### B. Q-value Table Updating and Controls Computation

In the RL framework, a learning agent interacts with a stochastic environment. Five key variables are exploited to model the interaction, wherein $S$ is the state variables set, $A$ is the control actions set, $P$ denotes the TPM for power demand, $R$ represents the reward function and $\mu \in (0, 1)$ means a discount factor.

Especially, these five variables are reified in the optimization control problem of energy management, such as $S = \{(SOC(t) | 0.3 \leq SOC(t) \leq 0.9\}$, $A = \{T_e(t) | 0 \leq T_e(t) \leq 900\}$, $r(s, a) \in R = \{\dot{m}_f(t) + \sigma(\Delta SOC)^2\}$ and $P$ is calculated in Section *II. D*.

The Q-value table in RL is defined as follows

$$Q(s,a) = r(s,a) + \mu \sum_{s' \in S} p_{s'a,s} Q(s',a') \quad (21)$$

where $p_{sa,s'}$ denotes the transition probability. $s'$ and $a'$ are the state variable and control action in the next step. This table is filled with state-action pair, i.e., Q($s$, $a$) and thus called a memory matrix. The memory of each state-action pair is applied to estimate the discounted sum of future rewards started from the current state and action policy.

Since the IMN value surpasses the threshold value, the memory matrix is decided to be updated. Before acquiring the new Q-value table, the TPMs and Q-value tables related to the primitive driving cycles need to be calculated and stored in the pre-learning process [40]. Assuming the number of primitive driving cycles is $N$, the existing TPMs and Q-value tables are denoted as

$$\{P_1, P_2, ..., P_N\} \text{ and } \{Q_1, Q_2, ..., Q_N\} \quad (22)$$

The current driving cycle information can be obtained via the onboard sensor in HEV and the current TPM of power demand is represented as $P_{new}$. The transferred coefficients for new Q-value table $Q_{new}$ related to the current driving cycle are determined as [32]

$$\delta_i = \frac{(T_f + \Delta IMN_{\max}) - IMN(P_i \| P_{new})}{\sum_{i=1}^{N}[(T_f + \Delta IMN_{\max}) - IMN(P_i \| P_{new})]}, i = 1...N \quad (23)$$
$$\Delta IMN_{\max} = \max_{i=1...N}[IMN(P_i \| P_{new})]$$

where $T_f \geq 0$ is the transfer factor and $\triangle IMN_{\max}$ is the maximum deviation of TPMs. A larger $T_f$ means that if the TPM of the sources cycle is closer to that of a new cycle, then more prior knowledge in the source Q-value table will be transferred into the new memory matrix.

Hence, the new Q-value table can be yielded by combining the transferred coefficients and primitive Q-value tables as

$$Q_{new} = \sum_{i=1}^{N} \delta_i \cdot Q_i \quad (24)$$

After acquiring the new Q-value table, the optimal RL-based controls can be derived in the Markov decision process (MDP) [41] in Matlab. The inputs are the TPMs, Q-value table and powertrain modeling and the outputs are the control policy and mean discrepancy of the memory matrix. The parameters of the calculation platform in this article are 2.7 GHz (CPU) and 3.8 GB (Memory). The adopted method is the Q-learning algorithm, whose pseudo-code is displayed in Table II.

TABLE II
ITERATION PROCESS OF RL-BASED CONTROL

| Method: Q-learning Algorithm |
|---|
| 1. Give a value for *K* and Q(*s*, *a*) |
| 2. Repeat step by step (*k*=1, 2, 3…) |
| 3. Using *ε*-greedy policy to select control action *a* |
| 4. Compute *r*, *s'* based on *s* and *a* |
| 5. Compare and decide *a\**=arg min$_a$ Q(*s'*, *a*) |
| 6. Q(*s*, *a*)←Q(*s*, *a*)+ *τ*(*r*(*s*, *a*)+ *μ*min $_{a'}$ Q(*s'*, *a'*)-Q(*s*, *a*)) |
| 7. Move to next step, *s*←*s'* |
| 8. finish when *s* is the terminal |

The discount factor $\mu$ is taken as 0.96 and the decaying factor $\tau$ is equal to $1/\sqrt{k+2}$ to accelerate the convergence rate. The discrete time step is 1 second and the iterative times *K* is 10000. The proposed adaptive energy management strategy is designed using the Q-value table transformation and Q-learning algorithm. The merits of the proposed transfer RL controller are discussed in the next section.

## IV. SIMULATION RESULTS AND DISCUSSION

The presented transfer RL-enabled adaptive and model-free optimal control policy is verified on the parallel hybrid powertrain in this section. First, the driving cycles transformation described in *Section II* is evaluated. The merits of the transformation are evaluated by comparing the reward function in RL. Furthermore, the influences of IMN threshold values on control performance are examined by comparing the fuel consumption based on the primitive and transferred Q-value tables. Numerical tests illuminate that the balance between control effectiveness and real-time performance can be tuned by the IMN value. Finally, the presented adaptive optimal control policy is compared with the conventional RL-based control and benchmarking DP to evaluate its availability and adaptability. Results imply that the presented control policy exceeds the conventional RL approach in both calculation speed and control performance.

### A. Merits of Transformation

An existing natural driving cycle is transformed into an equivalent one with the same characterizing measures (*α*, *β*, *γ*) of the real-time driving cycles. The resulting driving cycles from transformation are depicted in Fig. 4. Based on the expression (5), the original values of characterizing measures for the natural driving cycle are (*α*, *β*, *γ*) = (5039, 3.28, 0.54), and the target values in driving cycles A and B are (*α´*, *β´*, *γ´*) =

(4867, 3.07, -0.56) and (α´, β´, γ´) = (3451, 3.12, -0.20), respectively.

The dashed line represents the primitive driving cycle and the solid lines denote the transferred driving cycles. The zoom-in plots in Fig. 4 indicate that the transferred driving cycle for cycle A is different from that for cycle B. This attributes to the different target MTF components, which illustrates that the existing database can be converted into different types to imitate the real-time driving conditions.

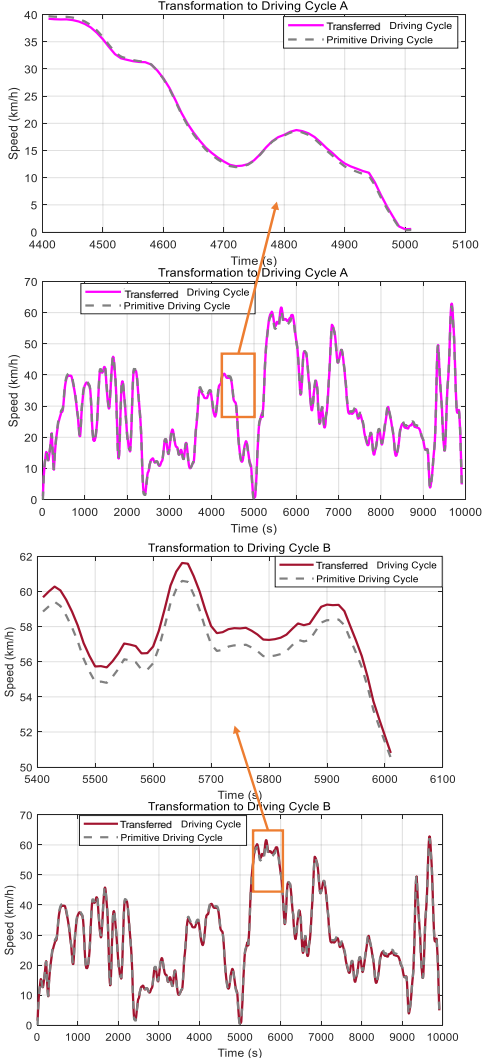

Fig. 4. Primitive and Transferred driving cycles related to A and B.

The transferred Q-value tables can be computed based on the TPM of power demand. The iteration process of the reward function based on the primitive and transferred Q-value tables is shown in Fig. 5. The merits of transformation including jumpstart, time to the threshold and asymptotic performance [42] are depicted in this figure. It is obvious that the transfer case is better than the non-transfer case in calculation speed and control performance, which demonstrates the necessity of transformation in RL framework.

### B. Influence of IMN Value

Based on the transferred Q-value table, the relevant RL-based control policy can be calculated. IMN values are exploited to quantify the differences between two TPMs and its threshold is used to decide the updating of the control policy.

Based on the driving cycles in Fig. 4, the relevant two TPMs are compared every 1000 second at different speed grades. As the IMN values exceed the threshold, the relevant control policies will be updated. Different IMN values result in various control policies, and they are utilized in driving cycle A and B in Fig. 4. Fig. 6 shows the SOC trajectories for different IMN values, in which the threshold value takes 0.1 and 0.3. It is noticed that the SOC trajectories are dramatically different for driving cycle A and but almost the same for driving cycle B. This is caused by the different controls that are computed via the transferred Q-value tables.

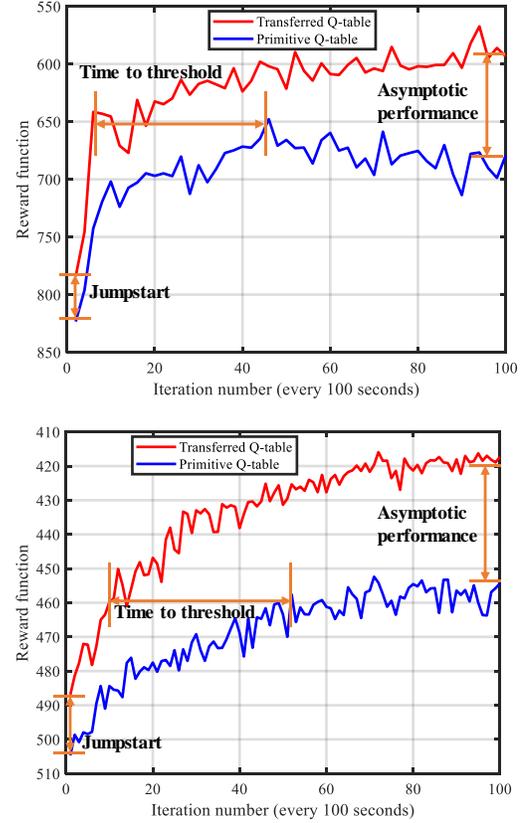

Fig. 5. Iteration process of the reward function for different Q-value tables.

The IMN values exceed the threshold means that the current driving cycle is very different from the historical ones, as described in Fig. 7. It implies that the existing controls are not suitable for the current driving conditions. Thus, the control policy based on the transferred Q-value table would replace the old one. From Fig. 7, it is obvious that control policy changes different times in driving cycle A for different IMN values and change same times in driving cycle B.

The number of updating times, fuel consumption and computation time of control policy are depicted in Table III. For the same IMN threshold, different driving cycles may experience various updating times. For the same driving cycle, different IMN thresholds may also experience various updating times. Furthermore, Table III indicates that lower IMN threshold value leads to more frequent control policy updating and more fuel consumption improving, nevertheless, the computation burden is heavier. Hence, IMN thresholds need to be tuned appropriately for online application. To compare the performance of different control policies in the next section, the IMN is chosen as 0.2 when considering calculation time and control performance together.

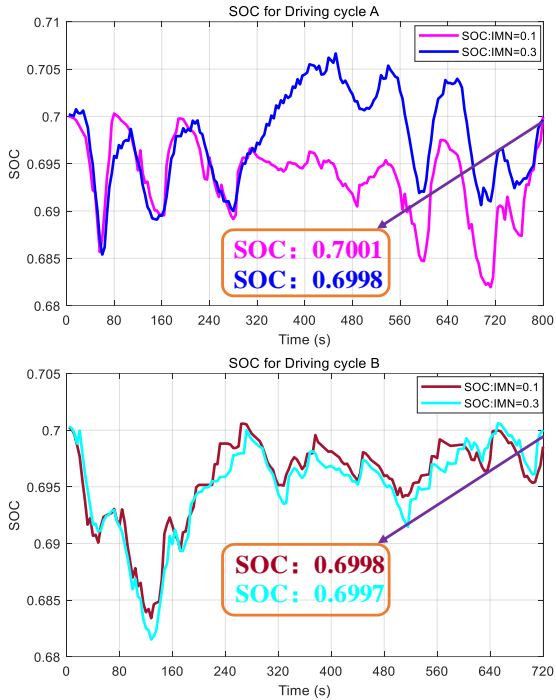

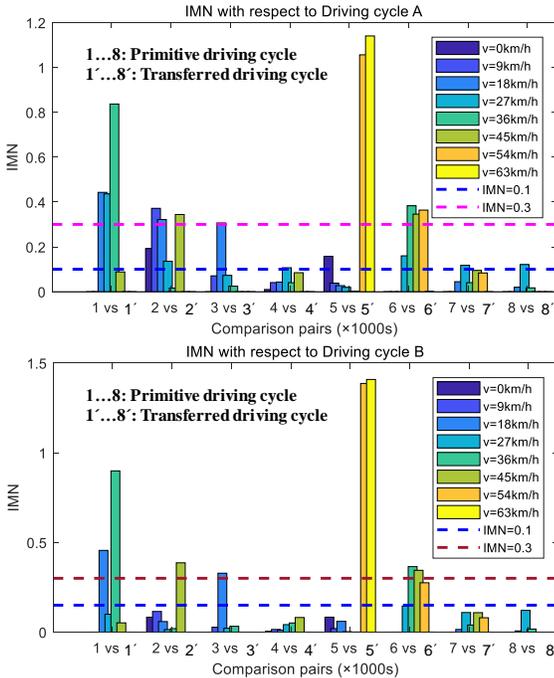

Fig. 6. SOC trajectories for driving cycles A and B with different IMN values.

Fig. 7. IMN values with time intervals 1000 seconds to cycle A and B.

TABLE III
THE UPDATED TIMES OF IMN AND FUEL CONSUMPTION FOR TWO CYCLES

| IMN | Updated Times$^A$ | Fuel (g)$^A$ | Time (s)$^A$ | Updated Times$^B$ | Fuel (g)$^B$ | Time (s)$^B$ |
|---|---|---|---|---|---|---|
| 0.1 | 8 | 583.3 | 120 | 5 | 428.29 | 74 |
| 0.3 | 4 | 661 | 57 | 5 | 440.74 | 75 |

$^A$ and $^B$ means driving cycles A and B.

## C. Comparison of Three Control Policies

The proposed transfer RL-based method is compared with the conventional RL and benchmarking DP on several driving cycles to certify its optimality and adaptability. Taking a real-time driving cycle C as an example, the state variable is SOC in battery and control variable is engine torque. The IMN threshold is 0.2 and the initial SOC is 0.70.

The SOC trajectories for the certain driving cycle C and the corresponding power split between the battery and engine are depicted in Fig. 8. It is obvious that the SOC trajectory based on the proposed control policy is close to that of DP-based control policy and clearly differs from that of the conventional RL control policy, which demonstrates its optimality. Fig. 8 shows the analogous variation in the power split results.

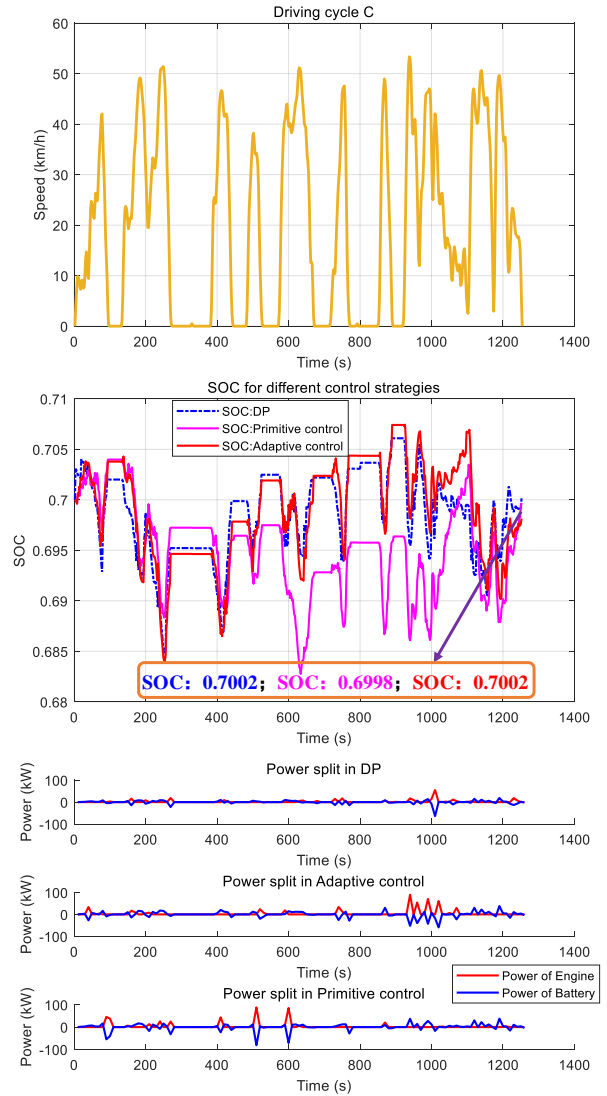

Fig. 8. Driving cycle C, SOC trajectories and power split for different control strategies.

The above observations are influenced by the updating of control policy that is decided by IMN thresholds. The IMN values exceed the threshold 0.2 four times, at 2000, 3000, 4000 and 7000 seconds, as shown in Fig. 9. Thus, the control policy updates at these time instants to adapt to the real-time driving conditions. Also, the convergence processes of the Q-value tables in the proposed controls and conventional RL-based controls are illustrated in Fig. 9. The mean discrepancy of action-value function in the proposed control is always lower than that in conventional RL control. It indicates that the proposed control is better than the conventional RL control in convergence rate. This discussion implies that the proposed energy management strategy adapts to the real-time driving conditions more suitable than the conventional RL control, which demonstrates its adaptability.

Also, Fig. 10 denotes the working conditions of engine corresponding to different control policies. Compared with the conventional RL-based control, more working points under the transfer RL-based and DP-based controls are located in the lower fuel consumption region, which results in the higher fuel economy.

the driving conditions via transferring the Q-value tables, which can lead to lower fuel consumption.

TABLE V
THE CALCULATION TIME FOR THREE CONTROL POLICIES

| Methods | Time[a] (s) | Relative increase (%) |
|---|---|---|
| Transfer RL | 59 | — |
| Conventional RL | 67 | 13.56 |
| DP | 245 | 315.25 |

[a] A 2.7 GHz microprocessor with 3.8 GB RAM was used.

The corresponding calculation time of these three control policies is contrasted in Table V. Obviously, the proposed method is faster than the conventional RL and DP, which enables the proposed transfer RL-based adaptive and model-free control online optimization feasible.

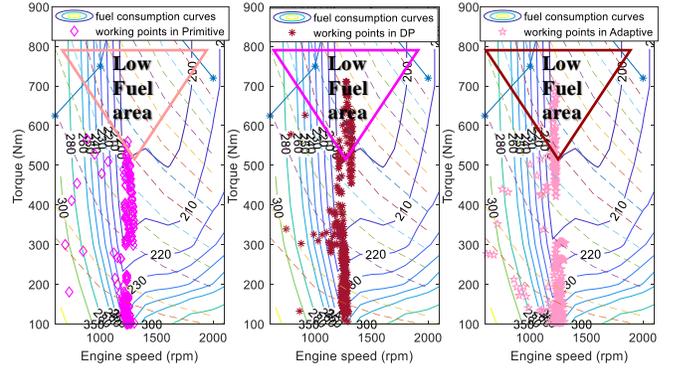

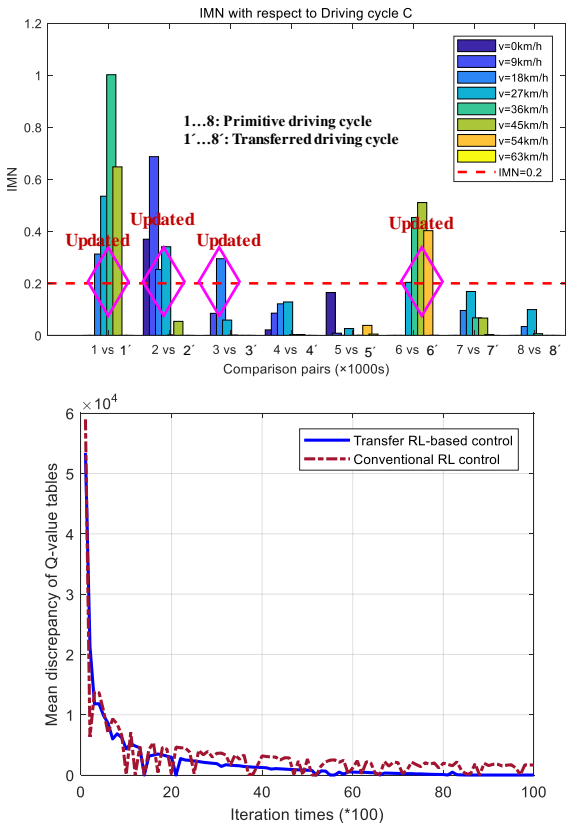

Fig. 9. IMN values with time intervals is 1000 second and the convergence processes of Q values.

TABLE IV
THE FUEL CONSUMPTION OF THREE CONTROL POLICIES

| Methods | Fuel consumption (g) | Relative increase (%) |
|---|---|---|
| DP | 1404.8 | — |
| Transfer RL | 1427.0 | 1.58 |
| Conventional RL | 1545.4 | 10.01 |

[a] A 2.7 GHz microprocessor with 3.8 GB RAM was used.

The different fuel consumption related to three control policies is denoted in Table IV. The fuel consumption of DP is lowest, and that of transfer RL is close to DP. The fuel consumption of conventional RL is highest, which is 10. 01% higher than transfer RL and 8.43% higher than the proposed method. In transfer RL, the control policy could change with

Fig. 10. Engine working points for different control strategies.

## V. CONCLUSION

This article proposes an adaptive energy management strategy for a hybrid electric vehicle with parallel topology using transfer reinforcement learning (RL) method. First, the up-level figures out how to transform the Q-value tables in RL framework via driving cycles transformation (DCT). This transformation converts the existent driving cycles database into an equivalent one, and the TPM of power demand is calculated to decide the Q-value table updating. Second, the induced matrix norm (IMN) is considered as a conclusive criterion to distinguish the differences of TPMs and to determine the transferred coefficients of Q-value table. Third, the lower-level determines how to establish the corresponding adaptive energy management strategy with the transferred Q-value tables using Q-learning algorithm.

The advantages in transformation and fuel economy are demonstrated by simulation results. Furthermore, the merits in control performance and calculation speed denote that the proposed transfer RL-based controller is able to be used in real-time situations.

The proposed transfer RL approach is indeed a simplified specification of the so called parallel learning [43] which aims to build a more general framework for data-driven intelligent control. The above testing results not only sheds light on real-time adaptive control design for online fuel economy improvement, but also indicates the potential of parallel learning in many other fields requiring flexible model-free control.